# New highly-anisotropic Rh-based Heusler compound for magnetic recording

Yangkun He[1,*], Gerhard H. Fecher[1,*], Chenguang Fu[1], Yu Pan[1], Kaustuv Manna[1], Johannes Kroder[1], Ajay Jha[2], Xiao Wang[1], Zhiwei Hu[1], Stefano Agrestini[3], Javier Herrero-Martń[4], Manuel Valvidares[4], Yurii Skourski[5], Walter Schnelle[1], Plamen Stamenov[2], Horst Borrmann[1], Liu Hao Tjeng[1], Rudolf Schaefer[6,7], S. S. P. Parkin[8], J. M. D. Coey[2] and Claudia Felser[1]

[1] *Max-Planck-Institute for Chemical Physics of Solids, D-01187 Dresden, Germany*
[2] *School of Physics, Trinity College, Dublin 2, Ireland*
[3] *Diamond Light Source, Harwell Science and Innovation Campus, Didcot, OX11 0DE, UK*
[4] *ALBA Synchrotron Light Source, Cerdanyola del Valles, 08290 Barcelona, Catalonia, Spain*
[5] *Dresden High Magnetic Field Laboratory (HLD-EMFL), Helmholtz-zentrum Dresden–Rossendorf, 01328 Dresden, Germany*
[6] *Leibniz Institute for Solid State and Materials Research (IFW) Dresden, Helmholtz strasse 20, D-01069 Dresden, Germany*
[7] *Institute for Materials Science, TU Dresden, D-01062 Dresden, Germany*
[8] *Max Planck Institute of Microstructure Physics, Halle, Germany*

**The development of high-density magnetic recording media is limited by the superparamagnetism in very small ferromagnetic crystals. Hard magnetic materials with strong perpendicular anisotropy offer stability and high recording density. To overcome the difficulty of writing media with a large coercivity, heat assisted magnetic recording (HAMR) has been developed, rapidly heating the media to the Curie temperature $T_c$ before writing, followed by rapid cooling. Requirements are a suitable $T_c$, coupled with anisotropic thermal conductivity and hard magnetic properties. Here we introduce $Rh_2CoSb$ as a new hard magnet with potential for thin film magnetic recording. A magnetocrystalline anisotropy of 3.6 MJm$^{-3}$ is combined with a saturation magnetization of $\mu_0 M_s$ = 0.52 T at 2 K (2.2 MJm$^{-3}$ and 0.44 T at room-temperature). The magnetic hardness parameter of 3.7 at room temperature is the highest observed for any rare-earth free hard magnet. The anisotropy is related to an unquenched orbital moment of 0.42 $\mu_B$ on Co, which is hybridized with neighbouring Rh atoms with a large spin-orbit interaction. Moreover, the pronounced temperature-dependence of the anisotropy that follows from its $T_c$ of 450 K, together with a high thermal conductivity of 20 Wm$^{-1}$K$^{-1}$, makes $Rh_2CoSb$ a**

**candidate for development for heat assisted writing with a recording density in excess of 10 Tb/in$^2$.**

The pace of doubling of information density on magnetic recording media has slackened in recent years, as the effective size of the perpendicularly recorded grains approached the superparamagnetic blocking diameter, which is the lower size limit for stable ferromagnetism, directly related to the magnetocrystalline anisotropy energy $K_1$. To resist demagnetization by random thermal fluctuation, the volume $V$ of a magnetic material must satisfy the empirical condition that $K_1V/k_BT > 60$ ($K_1V > 1.5$ eV), where $k_B$ and $T$ are the Boltzmann constant and ambient temperature, respectively[1-3]. For high-density magnetic recording media, strong perpendicular uniaxial anisotropy is required. To create a film for magnetic recording or a permanent magnet that remains fully magnetized regardless of its shape, $K_1$ should exceed $\mu_0 M_s^2$, where $M_s$ is the spontaneous saturation magnetization. Therefore, the magnetic hardness parameter $\kappa = \sqrt{K_1/(\mu_0 M_s^2)}$, a convenient figure of merit for permanent magnets[4], should be larger than 1. This is difficult to achieve in rare-earth free materials, other than CoPt or FePt with L1$_0$ structure.

The development of modern magnetic recording media spanned three generations. The first generation for tapes and discs depended on the shape anisotropy of acicular fine particles of ferrimagnetic or ferromagnetic oxides, $\gamma Fe_2O_3$ or $CrO_2$[5], which were magnetized in-plane. The second generation was based on hexagonal Co-Cr-Pt thin films with perpendicular magnetization, which is used in current hard discs[6]. But Cr decreases the $K_1$ of Co-Pt alloy, limiting the recording density. An ideal material for recording should be hard for storage and soft for writing, because of the limited fields that can be generated by the miniature writing electromagnet. L1$_0$ FePt, a hard magnet with $K_1$ = 6.6 MJm$^{-3}$ and $\kappa$ = 2.02, is the basis of new, third generation of heat-assisted magnetic recording (HAMR) media[7,8], where writing is realized by heating the material close to its Curie point with a laser-powered near-field transducer[9]. However, its high $T_c$ of 750 K makes rapid heating and cooling problematic and takes time[10]. To achieve a large temperature-variation in $K_1$, 10 % Cu is doped to obtain a 100 K reduction in $T_c$, but this is accompanied by a big drop of $K_1$ to 0.8 MJm$^{-3}$[3,11]. Additionally, unavoidable disordered A1 FePt impurities with a smaller $K_1$ and a lower $T_c$ make the stabilization of its properties difficult[12]. Therefore, it is useful to look for new materials with a strong $K_1$, a fairly suitable $T_c$ and good thermal conductivity for development as new thin film media

To achieve sufficient anisotropy, a non-cubic crystal structure (for example tetragonal or hexagonal) and a large spin-orbit interaction with the right sign is needed. When looking for new materials with uniaxial magnetocrystalline anisotropy, a significant deviation of the *c/a* ratio from 1 for tetragonal or $\sqrt{8/3}$ for hexagonal structures, is sought. Moreover, it is important to pair a 3*d* metal (Mn, Fe, Co) that provides a substantial magnetic moment with a heavy atom that enhances the spin-orbit interaction. Rh is a 4*d* element where the spin-orbit interaction is larger than in 3*d* elements, and it acquires a small induced magnetic moment when paired with a 3*d* metal. Therefore, it is worthwhile looking for new opportunities in magnetic recording among Rh-based alloys with a tetragonal or hexagonal structure.

Heusler alloys are a large family of compounds with formula $X_2YZ$, where X and Y are usually transition metals and Z is a main group element[13,14]. The abundant choice of elements provides great scope to search for new materials with specific properties. $Rh_2CoSb$ has a $c/\sqrt{2}a$ ratio of 1.24, the largest among all reported Rh-based magnetic Heusler compounds ($c/\sqrt{2}a$ is used to describe the tetragonal distortion relative to a cubic lattice). A magnetic moment per formula of 1.4 $\mu_B$ in an 0.7 T applied field and a $T_c$ of about 450 K were reported for polycrystalline samples many years ago[15]. Recent *ab-initio* calculations proposed uniaxial anisotropy with an easy *c*-axis[16]. Both experiments and calculations suggest that $Rh_2CoSb$ might be an interesting hard magnetic material.

Here we study the anisotropic magnetic, thermal and transport properties, as well as the temperature-dependent anisotropy of single crystals of $Rh_2CoSb$ and investigate the contribution of rhodium to the magnetism.

**Results**

**Crystal structure**

Magnetic Heusler alloys with a tetragonal structure are often Mn-based with space group $I\bar{4}m2$ [17-22]. However, the Mn-Mn magnetic coupling is usually antiferromagnetic at the nearest-neighbour separation in this structure, leading to ferrimagnetic order with relatively low magnetization (**Figure 1**a). Some tetragonal Rh-based Heusler alloys including $Rh_2Mn_{1.12}Sn_{0.88}$, $Rh_2FeSn$, and $Rh_2CoSn$ and $Rh_2CoSb$[15] with 3*d* transition elements Mn, Fe and Co, crystallize in a tetragonal structure with space group *I*4/mmm ($D0_{22}$ structure). The 4*d* element Rh provides a strong spin orbit interaction that contributes to the magnetocrystalline anisotropy in the tetragonal lattice[23].

Both XRD and TEM experiments show that $Rh_2CoSb$ has a well-ordered tetragonal $D0_{22}$ structure with $a$ = 4.0393 (6) Å and $c$ = 7.1052 (7) Å. Here, the 4$d$ (0 0 ¼) site is occupied by Rh, 2$b$ (0 0 ½) by Co, and 2$a$ (0 0 0) by Sb as shown in Figure 1c. The tetragonal distortion $c/\sqrt{2}a$ = 1.2436(3) agrees with the published value[15], being the largest among all reported Rh-based magnetic Heusler compounds. A sketch of the easy-axis energy surface is included in Figure 1c.

Unlike many Mn-based Heusler compounds, which are often cubic at high temperature and may undergo a martensitic phase transition to become tetragonal, $Rh_2CoSb$ does not have a first-order transition at high temperature and remains in the tetragonal phase at all temperatures up to the melting point of 1482 K. A differential scanning calorimeter (DSC) measurement showing only the melting transition is presented in Supplementary Information. Since the phase melts congruently, single crystals can be grown by the Bridgeman method. They were cut along the $c$ and $a$ axes for further measurements and fully characterized by x-ray diffraction (XRD), wavelength dispersive x-ray spectroscopy (WDX) and transmission electron microscopy (TEM) in Supplemental Information. WDX establishes a homogenous composition of $Rh_{50.3}Co_{25.6}Sb_{24.1}$ (see Supplement). Errors in composition range from 0.1 – 0.2 %.

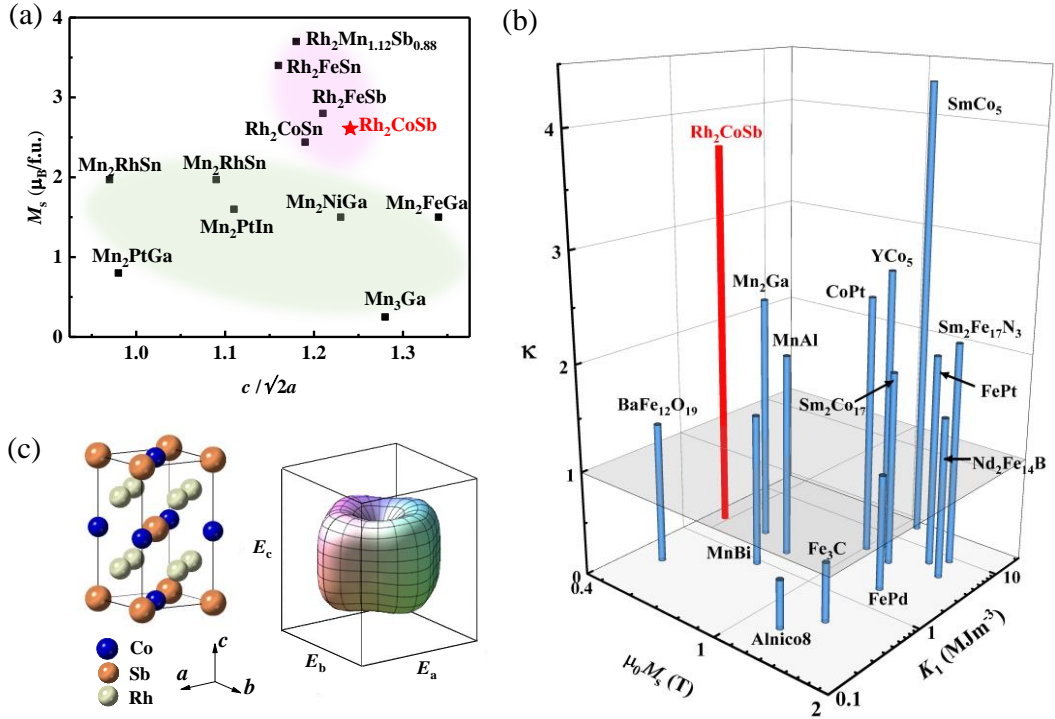

**Figure. 1.** Tetragonal structure of Rh$_2$CoSb. a, Tetragonal distortion versus magnetic moment for Mn-based and Rh-based Heusler compounds. Rh-based compounds show both large distortion and large magnetization. Note that some compounds do not saturate in the applied magnetic fields of 7 T for Mn$_2$PtGa, 10 T for Mn$_2$FeGa, 6.6 T for Rh$_2$CoSn and Rh$_2$FeSn, and 0.7 T for Rh$_2$FeSb and Rh$_2$Mn$_{1.12}$Sb$_{0.88}$. b, Comparison of the magnetic hardness parameter κ with other hard magnets. The light grey plane marks the threshold $\kappa = 1$. c, Unit cell of tetragonal Heusler compound Rh$_2$CoSb, whose magnetocrystalline anisotropy surface shows easy-axis anisotropy along *c*.

## Magnetic properties

Magnetic properties were measured along the *c* and *a* axes of single crystals as illustrated in **Figure 2**. A magneto-optical Kerr microscopy study[24] shows surface domains of a two-phase branched domain pattern of higher generation on the (001) surface, which is commonly observed in uniaxial magnets. At 2 K, the magnetization along $c$ saturates easily, at 37.5 Am$^2$kg$^{-1}$. Taking the density of 11,030 kg m$^{-3}$ into consideration, the saturation magnetization $\mu_0M_s$ is 0.52 T, which is similar to that of pure nickel. It corresponds to a moment of 2.6 $\mu_B$ per Rh$_2$CoSb formula unit. The non-integer magnetic moment per unit cell indicates that Rh$_2$CoSb is not a half metal. However, a field $\mu_0H_a$ of 17.5 T is required to saturate the sample along the $a$ axis, where $H_a$ is the anisotropy field. The magnetocrystalline anisotropy constant $K_1 = ½ \mu_0H_aM_s$ is 3.6 MJm$^{-3}$, which is larger than for any other rare-earth free compound except for CoPt and FePt. Its relatively small $M_s$ makes $\kappa$ larger than for these materials and coercivity may exceed that of FePt[3]. Both $\mu_0M_s$ and $K_1$ decrease with increasing temperature, as shown in Figure 2b, but at room temperature they are still 0.44 T and 2.2 MJm$^{-3}$, respectively. Big Barkhausen jumps observed during $c$ axis demagnetization, together with a high initial susceptibility after thermal demagnetization indicate that single-crystal Rh$_2$CoSb is a nucleation type magnet[25,26]. Detailed measurements of the magnetization along $c$ and $a$ at different temperatures are reported in the Supplemental information.

The Curie temperature is deduced to be 450 K from the temperature scans of magnetization in a 10 mT field. The $M$-$T$ curve measured along the hard axis in a magnetic field of 0.5 T exhibits a sharp peak at 440 K, indicating a rapid built-up of anisotropy just below $T_c$. The slope d$K_1$/d$T$ is ~ 20 kJm$^{-3}$K$^{-1}$ (twice as large as for FePt[27]). The ac-susceptibility $\chi$ near $T_c$ is shown in Figure 2d. It exhibits a typical Hopkinson peak along the $c$ axis[15,28]. Since $\chi \propto M_s^2/K_1$, during cooling the rate of increase of $M_s$ and $K_1$ vary at different temperatures near $T_c$, giving a trough in ac susceptibility below the Hopkinson peak. There is nothing unusual in our *M-T* and *M-H* measurements at this temperature.

We compare in Figure 1b the magnetic hardness parameter of $Rh_2CoSb$ with other materials that exhibit $\mu_0 M_s > 0.4$ T, which is taken as a threshold necessary for useful stray fields. $\kappa$ is a practical figure of merit for hard magnetic materials that must be be greater than 1 if the material is to resist self-demagnetization when fabricated into any desired shape[4]. $Rh_2CoSb$ has $\kappa = 4.1$ at 2 K and 3.7 at room temperature, which is more than any other rare-earth free magnet. Only $SmCo_5$ has a larger value.

All the indications are that $Rh_2CoSb$ has the potential to be a good hard magnet with strong uniaxial anisotropy. There is no structural transition, avoiding the problems of decomposition (MnBi decomposes at 628 K) or twinning (like many nearly-cubic rare-earth free magnets such as MnGa, MnAl, FePt and CoPt[29]). Unlike many rare-earth magnets, the sample is stable in air and its magnetic properties have been found to remain unchanged for a year.

Thanks to the strong $K_1$ at 300 K, $Rh_2CoSb$ promises a significant reduction in thermally stable grain size from diameter $D = 7–9$ nm in today's perpendicular Co-Cr-Pt media down to $D = 4–5$ nm in future $Rh_2CoSb$ media, resulting in a potential storage density of more than 10 Terabit/inch$^2$ (for $D = 5$ nm with half area occupied). Detailed calculation can be found in Supplemental information. HAMR media with grain diameters of only a few nm are hard to fabricate because the grains musts be exchange decoupled by an intergranular material. Powder XRD data revealed the presence of a few percent of a RhSb secondary phase in our polycrystalline samples. Nonmagnetic RhSb appears at the grain boundaries, pinning the domains and creating hysteresis. It is a candidate for separating the nanocrystal grains in $Rh_2CoSb$ thin film media.

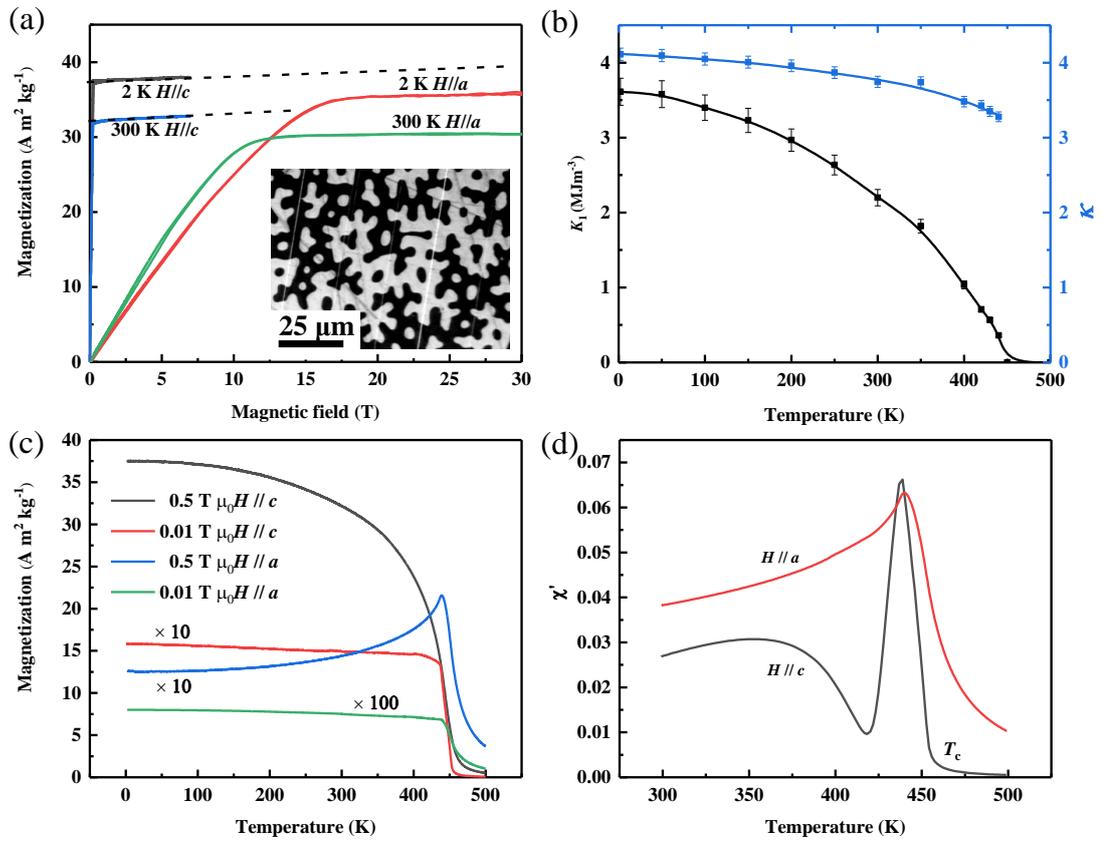

**Figure 2.** Magnetic properties of $Rh_2CoSb$. a, Magnetization curves at 2 K and 300 K with the field along the *a* or *c* axes. The insert shows a pattern of branched domains at the surface perpendicular to the *c* axis, typical of a strong uniaxial ferromagnet. b, Magnetocrystalline anisotropy calculated from magnetization curves at different temperatures. The blue line

shows the magnetic hardness parameter $\kappa$ at different temperatures. c, Magnetization at different temperatures with the field along *a* or *c* axes. For better visibility some values are enlarged by a factor of 10 or 100, depending on the field direction and the value of *H*. d, AC-susceptibility near $T_c$, showing a typical Hopkinson peak, due to the rapid increase of anisotropy just below $T_c$.

**Transport properties.**

Transport properties of $Rh_2CoSb$ are presented in **Figure 3**. The resistivity is very anisotropic. With increasing temperature, the *a*-axis resistivity increases monotonically from 53 μΩ cm at 2 K to 192 μΩ cm at $T_c$, after which the resistivity, dominated by spin disorder scattering, tends to saturate[30]. The *c*-axis resistivity is less than half as large at 2 K, 21 μΩ cm, but it increases with temperature and shows a similar trend to the *a*-axis resistivity. The difference is due to intrinsic mobility and extrinsic domain wall scattering perpendicular to *c* axis[31]. Magnetoresistance and Hall measurements are also very anisotropic (see Supplement).

The Seebeck coefficient is about -10 $WK^{-1}m^{-1}$ at 300 K along both axes (*c* and *a*) with an error of ±10%. The opposite signs of the Hall effect (positive) and Seebeck coefficient (negative) indicate the co-existence of both light holes and heavy electrons at the Fermi energy[32]. Detailed data are presented in the Supplement. The spin polarization *P* of the electrons at the Femi level was deduced from a point contact Andreev reflection measurement at 2 K (see Supplement). The measured value of *P* is 13 % and agrees with well with the calculated spin polarization of the density of states at the Fermi energy (see Supplement). The transport polarization will be different due to different effective masses of minority and majority electrons[33].

The measured thermal transport properties in Figures. 3b and 3c along *c* and *a* axes are also highly anisotropic. The total thermal conductivity along *c* is about twice as large as along *a*, and it is mainly explained by the carrier contribution, following the Wiedemann–Franz law[34]. The remaining, almost isotropic, part is dominated by the phonon contribution. The slight upturn of the thermal conductivity at temperatures above 200 K may be due to uncertainty in the radiative heat losses, which is estimated to be about ±10%. In addition, magnons or electron-magnon interactions might influence the thermal conductivity. The *limiting* Lorenz number generally depends a little on temperature[34]. The anisotropy reflects the anisotropic electronic structure, which is the origin of the giant magnetocrystalline anisotropy. The *c*-axis thermal conductivity of 20 Wm$^{-1}$K$^{-1}$ at room temperature is roughly twice that of unsegregated L1$_0$ FePt (11-13 Wm$^{-1}$K$^{-1}$ [35,36]), or A1 FePt (9 Wm$^{-1}$K$^{-1}$ [36]). The anisotropic transport properties, including magnetoresistance, anomalous Hall effect, and Seebeck effect, result from the anisotropic electronic structure, that leads to an anisotropic mobility of the charge carriers.

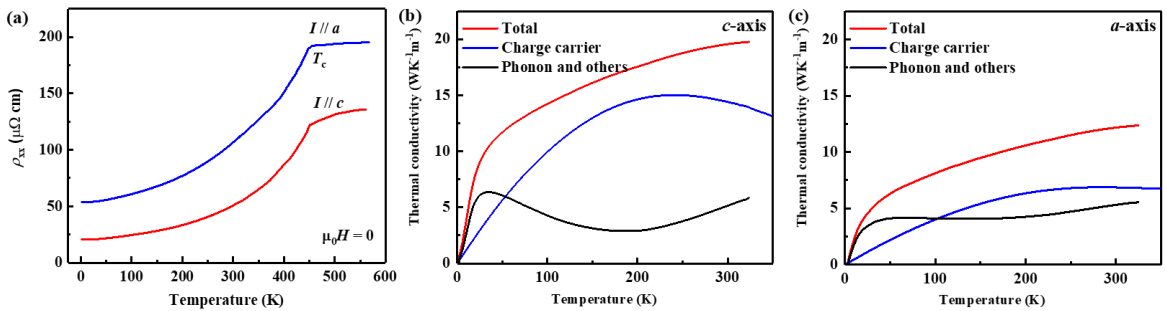

**Figure. 3.** Transport properties of Rh$_2$CoSb. a, Longitudinal resistivity along the *c* (red curve) and the *a* axis (blue curve). b and c, Thermal conductivity along the *c* and *a* axes and the charge carrier contribution (estimated from the Wiedemann–Franz law as $LT/\rho_{xx}$, where $L$ = 2.44 WΩK$^{-2}$ is the Lorenz number). The remaining part is mainly the phonon contribution.

**Discussion**

Little information is available about the magnetism of hard magnetic 3$d$-4$d$ intermetallic compounds, other than FePd, which has the tetragonal L1$_0$ structure with $K_1$ = 1.8 MJm$^{-3}$, and YCo$_5$ which has the hexagonal CaCu$_5$ structure[5] with $K_1$ = 6.5 MJm$^{-3}$. A number of ternary, tetragonal Rh-based intermetallics are known to order ferromagnetically with a Curie point above room temperature[15]. Only one atom out of four in the Rh$_2$CoSb formula is cobalt, but the high $T_c$ of 450 K indicates a strong exchange interaction. We see from our *ab-initio* calculations (Supplemental information) that Rh is clearly in a spin-polarized state, and that it contributes to the ferromagnetism. The measured magnetic moment per formula of 2.6 μ$_B$ is much higher than the ordinary ~1.6 μ$_B$ moment of Co in the elemental state or in metallic alloys with other 3$d$ elements. The difference should be attributed, at least partially, to Rh, or else to enhanced Co spin and orbital moments, as there is no magnetic contribution from Sb.

We also prepared polycrystalline Rh$_2$FeSb with the same crystal structure, which exhibits easy-plane magnetization. At 300 K under 7 T, the incompletely saturated moment reaches 3.8 μ$_B$ per formula, far greater than the 2.2 μ$_B$ of metallic iron. The Curie temperature of 510 K is even higher than that of Rh$_2$CoSb, which is in contrast with other isostructural Fe and Co intermetallics. A detailed comparison of the spin and orbital contributions for Rh$_2$FeSb and Rh$_2$CoSb deduced from *ab-initio* calculations is provided in the Supplemental Information. All the evidence illustrates that Rh plays an important role in the ferromagnetism of these ternary compounds. In fact, it has already been shown to carry an induced moment in binary Fe-Rh and Co-Rh alloys that is roughly one third of the 3*d* moment[37,38].

To reveal the site specific magnetic moments of Rh and Co, X-ray magnetic circular dichroism (XMCD) measurements were performed at the *L*$_{2,3}$ edges of Co and Rh, respectively. The spin and orbital moments of each element[39-41] are determined from XMCD using sum rule analysis[42,43]. The same number of 7.8 electrons in the valence *d* shell is assumed for both Co and Rh. This value was found in fully relativistic *ab-initio* calculations. The measured XMCD spectra are shown in Figure 4. The XMCD signal has the same sign in the spectra obtained at the Co and Rh *L*$_{2,3}$ edges, which proves that the coupling between Co and Rh is ferromagnetic.

The XMCD signal at the *L*$_3$ edge for Co is considerably larger than that at the *L*$_2$ edge, which indicates a sizeable orbital contribution. Sum rule analysis reveals that spin and orbital moments for Co are 1.53 ±0.15 μ$_B$ and 0.42 ±0.04 μ$_B$, respectively. The orbital moment of Co in Rh$_2$CoSb far surpasses that of elemental Co, where the value is 0.15 μ$_B$[44]. Therefore, the orbital moment of Co makes a sizeable contribution to the overall magnetization of 1.95 ±0.19 μ$_B$. The presence of a large orbital moment is also evident in Figure 2a from the 0.2 μ$_B$ difference of saturation magnetization between the *c* and *a*.

The spin and orbital moments of Rh obtained from the sum rule analysis are 0.28±0.03 μ$_B$ and 0.020±0.002 μ$_B$, respectively, resulting in a total moment of 0.30±0.03 μ$_B$. These values account for the reduced photon polarization at high energies. Using these moments, together with those of Co, the total magnetic moment of Rh$_2$CoSb should amount to 2.55±0.22 μ$_B$, which is in good agreement with the magnetic measurement of 2.6 μ$_B$.

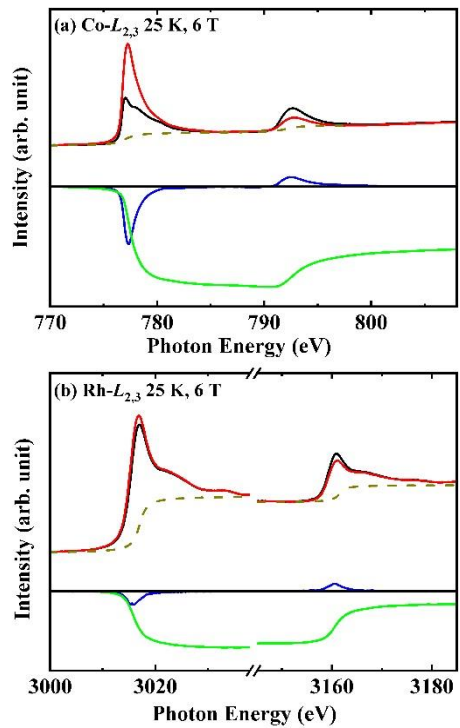

**Figure. 4. XAS and XMCD spectra of Rh$_2$CoSb. a,** Co $L_{2,3}$ and **b,** Rh $L_{2,3}$ XAS and XMCD spectra. Spectra were taken at 25 K temperature and 6 T induction field. The photon helicity is oriented parallel ($\mu^+$) in black solid line or antiparallel ($\mu^-$) in red solid line to the magnetic field. The background is shown in the black dash line with edge jumps.

The Co atoms have a substantial orbital magnetic moment in Rh$_2$CoSb, both from experiments and *ab-initio* calculations. The orbital moment was 27% of the spin moment, compared to 5% in elemental α-Co (hcp) and 3% for Fe in calculations of the sister compound Rh$_2$FeSb. The larger orbital moment for cobalt reflects the hybridization of Co 3*d* states with 4*d* states of Rh, which has a stronger spin-orbit interaction than Co.

In many Heusler compounds, the martensitic transition from a cubic to a tetragonal structure is explained by a band Jahn-Teller effect[23] that results in a splitting of energy levels by a modification of their width. We also calculated Rh$_2$CoSb in the higher energy cubic $L2_1$ structure, but saw no sign of Jahn-Teller type splitting in the electronic structure. The formation enthalpies for our compound in the tetragonal *I*4/*mmm* (space group 139), inverse tetragonal $I\bar{4}m2$ (space group 119) and cubic $L2_1$ (space group 225) structures are -753, -325 and -221 meV, respectively. The energy difference between the cubic and tetragonal structures is 532 meV, significantly more than in other Rh-based Heusler compounds. Thus, the hypothetical cubic to tetragonal phase transition temperature ($T_{\text{cub-tet}}$) should be above the melting point. Among the reported Rh$_2$CoZ Heusler compounds, only Rh$_2$CoSb and Rh$_2$CoSn are tetragonal at room temperature; the others are cubic. However, Rh$_2$CoSn has a martensitic transition at around 600 K with a cubic structure at higher temperature[23]. By avoiding any such transition, Rh$_2$CoSb is superior to other Rh$_2$CoZ alloys for HAMR media.

Our work will help to design new rare-earth free materials with strong uniaxial magnetocrystalline anisotropy, for which we need:

1) a low-symmetry crystal structure (tetragonal or hexagonal);

2) heavy atoms with a large spin-orbit interaction (4*d*, 5*d* or 5*f* elements);

3) another possible element with the right electronegativity to help stabilize the

structure and help produce an advantageous electronic structure.

The magnetic moments of heavy non-rare-earth elements are small at best. Therefore, the design rule for rare-earth free metallic magnets is to pair a 3$d$ metal (Mn, Fe, Co) that can provide most of the magnetization with a heavy atom that supplies a large spin-orbit interaction. Rh, Ir, Pd and Pt are all possible choices, since they are in the same group as Co or Ni and they may exhibit a significant induced magnetization[37,38,45]. In fact, Pd is already ferromagnetic with $T_c$ = 17 K if lightly doped with 0.5% of Co[46] and when it is alloyed with 50% Co, it has the highest $T_c$ among the binary compounds CoRh (550 K), CoIr (40 K), CoPd (950 K) and CoPt (840 K)[47,48]. Our study outlines the design principles for new rare-earth free hard magnets, but it is unrealistic to imagine that Rh-based alloys will ever be used as bulk functional materials, in view of its high cost. This is much less important in functional thin films, because the quantities used are tiny. Strong perpendicular anisotropy is of increasing importance in spintronics and especially in HAMR; there thermal conductivity is also an important consideration.

Ref. 16 reported the preparation of Rh$_2$CoSb films by sputtering or ion-beam deposition between 373-873 K. A large perpendicular anisotropy was observed. Disorder is common in sputtered films. Our experiments indicate that the Rh is well ordered, but there might be some disorder between Co and Sb atoms. From *ab-initio* calculations, the most stable crystal structure for both ordered and disordered phases is the tetragonal D0$_{22}$ structure and the lattice constant is very similar. The total magnetic moment in the fully Co-Sb disordered state is about 20% larger compared to the completely ordered state. The orbital moment is nearly constant and independent of the degree of disorder, so the increase of the total moment is attributed to the spin moment.

We compare $Rh_2CoSb$ with $L1_0$ FePt[5] for HAMR in Table 1. The magnetic properties for $Rh_2CoSb$ at 2 K are shown with brackets, those at 300 K without. The magnetic properties for $L1_0$ FePt do not vary much between 2 K and 300 K. The writing speed is limited by the cooling time when the total heat produced by laser transfers to the substrate. Since $Rh_2CoSb$ has a much lower Curie point and higher thermal conductivity, its writing speed will be roughly 6.8 times faster than $L1_0$ FePt (see Supplementary information).

**Table 1.** Comparison of $Rh_2CoSb$ with $L1_0$ FePt for HAMR at room temperature. The properties for $Rh_2CoSba$ at 2 K is shown in bracket.

| Materials | Crystal structure | $K_1$ (MJm$^{-3}$) | $\mu_0 M_s$ (T) | $\mu_0 H_a$ (T) | $\kappa$ | $T_c$ (K) | $T_{cub-tet}$ | Thermal stability | Thermal conductivity (Wm$^{-1}$K$^{-1}$) | Writing speed |
|---|---|---|---|---|---|---|---|---|---|---|
| $Rh_2CoSb$ | $I4/mmm$ | 2.2 (3.6) | 0.44 (0.52) | 12 (17) | 3.7 (4.1) | 450 | > melting point | Stable | 20 in $c$ axis 12 in $a$ axis | Fast |
| $L1_0$ FePt | $P4/mmm$ | 6.6 | 1.43 | 11 | 2 | 750 | 1573 K | Transition to A1 FePt | 11 | Slow |

**Conclusion**

In summary, $Rh_2CoSb$ is a uniaxial ferromagnet with a remarkable magnetocrystalline anisotropy of 3.6 MJm$^{-3}$, due to a large unquenched orbital moment of 0.42 $\mu_B$ on Co that arises from hybridization with the surrounding Rh, where spin orbit coupling is strong. The magnetic hardness parameter of $\kappa = 3.7$ at room temperature is the highest observed so far in any rare-earth free magnet. The relatively low Curie point of $T_c$ = 450 K leads to a large temperature-dependence of $K_1$ from a high base at room temperature, which is an asset for data writing. The anisotropic thermal conductivity, especial its large $c$ axis value of 20 Wm$^{-1}$K$^{-1}$, which is much larger than that of the current FePt HAMR material, is important for cooling, which would lead to a 6.8 times faster writing speed than FePt. Unlike FePt with its order/disorder phase transition, Rh$_2$CoSb is a stable phase without any structural transition below the melting point and its properties are stable in air. All these features commend Rh$_2$CoSb as a candidate for HAMR media with a recording density of more than 10 Tb/in$^2$ and high writing speed.

**Methods**

**Single crystal growth.** The single crystals of Rh$_2$CoSb were grown by the Bridgeman method. First, the high purity (> 99.99%) elements Rh, Co and Sb were cut into small pieces and arc-melted together to prepare polycrystalline samples. The initial atomic ratio of Rh, Co and Sb was 2:1:1.03. Additional Sb was added due to compensate for its high vapor pressure. From powder XRD data, there were always a few percent of a RhSb secondary phase in the polycrystalline samples. Hence the polycrystalline materials were heated up to 1600 K (above the melting point of RhSb at 1583 K) for 3 days to achieve a homogeneous liquid before beginning the single crystal growth, which took about 5 days during cool down to 1273 K. The composition of the crystals was checked by wavelength dispersive X-ray spectroscopy that showed a homogenous composition of $Rh_{50.3}Co_{25.6}Sb_{24.1}$. The crystals were characterized by powder X-ray diffraction as single-phase with a tetragonal structure. The orientation of the single crystals was confirmed by the Laue method.

**Magnetization measurements.** These were conducted on single crystals with the magnetic field applied along either the *a* or *c* axis using a vibrating sample magnetometer (MPMS 3, Quantum Design). The sample size was 3.80×0.78×1.30 mm$^3$. For measurement with the field along the *a* axis, the sample was carefully immobilized with glue in view of the strong torque. High-field magnetization measurements were performed in the Dresden High Magnetic Field Laboratory using a pulsed magnet.

**Magneto-optical Kerr microscopy.** Domain images were performed by using the polar Kerr effect in a wide-field magneto-optical Kerr microscope at room temperature with a polished 1 mm thick single crystal with (001) surface. Detailed description of the method can be found elsewhere[49].

**Electrical transport measurements.** The longitudinal and Hall resistivities were measured on a Quantum Design PPMS 9 using the low-frequency alternating current (ACT) option for data below 320 K. The longitudinal resistivity was measured with standard four-probe method, while for the Hall resistivity measurements, the five-probe method was used with a balance protection meter to eliminate possible magnetoresistance signals[50]. Longitudinal resistivity measurement above 320 K, were measured on a home-made device using a four-probe method and careful calibration. The accuracy of resistivity measurement is ±5%.

**Thermal transport measurements.** The thermal conductivity and Seebeck thermopower were measured adiabatically in the Quantum Design PPMS using thermal transport option (TTO). The uncertainty of the radiative heat losses at high temperature and the uncertainty of the geometry is estimated as ±10%.

**Andreev reflection measurements.** Measurements are performed on a polished (001) crystal surface in a flow of helium vapor, using a mechanically sharpened Nb tip in the absence of an external magnetic field. Data are analyzed using the modified BTK model, as detailed elsewhere[51]. The best fit to spectrum is obtained with barrier parameter $Z\sim$ of 0.35, an electron temperature of 2 K and a spin polarization of 13%.

**XMCD measurements.** X-ray magnetic circular dichroism (XMCD) was measured at the beamline BL29 (BOREAS) of the synchrotron ALBA in Barcelona (Spain). XMCD spectra at the $L_{2,3}$ absorption edges of Co and Rh were taken at a temperature of 25 K in a vacuum chamber with a pressure of $10^{-9}$ mbar. The x-ray absorption spectra (XAS) were measured using circular polarized light with photon helicity parallel ($\mu^+$) or antiparallel ($\mu^-$) to the fixed magnetic field in the sequence $\mu^+\mu^-\mu^-\mu^+\mu^-\mu^+\mu^+\mu^-$ to disentangle the XMCD. An induction field of 6 T was applied along the $c$ axis. The polarization delivered by the Apple II-type elliptical undulator was close to 100% for the Co $L_{2,3}$ edges and 70% for the Rh $L_{2,3}$ edges. The spectra were recorded using the total yield mode.

**Supporting Information**

Supporting Information is available from the Wiley Online Library or from the author.

**Acknowledgement**

This work was financially supported by the European Research Council Advanced Grant (No. 742068) "TOPMAT", the European Union's Horizon 2020 research and innovation programme (No. 824123) "SKYTOP", the European Union's Horizon 2020 research and innovation programme (No. 766566) "ASPIN", the Deutsche Forschungsgemeinschaft (Project-ID 258499086) "SFB 1143", the Deutsche Forschungsgemeinschaft (Project-ID FE 633/30-1) "SPP Skyrmions", the DFG through the Würzburg-Dresden Cluster of Excellence on Complexity and Topology in Quantum Matter ct.qmat (EXC 2147, Project-ID 39085490) and the DFG through SFB 1143. We acknowledge the support of the High Magnetic Field Laboratory Dresden (HLD) at HZDR, members of the European Magnetic Field Laboratory (EMFL).

**Conflict of Interest**

The authors declare no conflict of interest.

**Author contributions**

Single crystals were grown by Y.H. and K.M. The characterization of the crystal, magnetic and transport measurement were performed by Y.H. with the help of C.Fu, Y.P., J.K., A.J., Y.S.,W.S., P.S., H.B., R.S. and S.S.P.P. XMCD was measured by X.W., Z.H., S.A., J.H., M.V. and L.H.T. First principle calculations were carried out by G.H.F. All the authors discussed the results. The paper was written by Y.H., J.M.D.C. and G.H.F. with feedback from all the authors. The project was supervised by C. Felser.

# New highly-anisotropic Rh-based Heusler compound for magnetic recording

# Supplementary Information

Yangkun He[1,*], Gerhard H. Fecher[1,*], Chenguang Fu[1], Yu Pan[1], Kaustuv Manna[1], Johannes Kroder[1], Ajay Jha[2], Xiao Wang[1], Zhiwei Hu[1], Stefano Agrestini[3], Javier Herrero-Martń[4], Manuel Valvidares[4], Yurii Skourski[5], Walter Schnelle[1], Plamen Stamenov[2], Horst Borrmann[1], Liu Hao Tjeng[1], Rudolf Schaefer[6,7], S. S. P. Parkin[8], J. M. D. Coey[2] and Claudia Felser[1]

[1]*Max-Planck-Institute for Chemical Physics of Solids, D-01187 Dresden, Germany*
[2]*School of Physics, Trinity College, Dublin 2, Ireland*
[3] *Diamond Light Source, Harwell Science and Innovation Campus, Didcot, OX11 0DE, UK*
[4]*ALBA Synchrotron Light Source, Cerdanyola del Valles, 08290 Barcelona, Catalonia, Spain*
[5] *Dresden High Magnetic Field Laboratory (HLD-EMFL), Helmholtz-zentrum Dresden–Rossendorf, 01328 Dresden, Germany*
[6]*Leibniz Institute for Solid State and Materials Research (IFW) Dresden, Helmholtz strasse 20, D-01069 Dresden, Germany*
[7]*Institute for Materials Science, TU Dresden, D-01062 Dresden, Germany*
[8]*Max Planck Institute of Microstructure Physics, Halle, Germany*

## 1. Single crystal growth and composition

The composition of $Rh_2CoSb$ single crystal has been identified by both energy-dispersive X-ray spectroscopy (EDX) and wave length dispersive X-ray spectroscopy (WDX), which shows a homogenous composition. The average values and standard deviation are shown in the Table S1. The sample has a homogenous composition of $Rh_{50.3}Co_{25.6}Sb_{24.1}$.

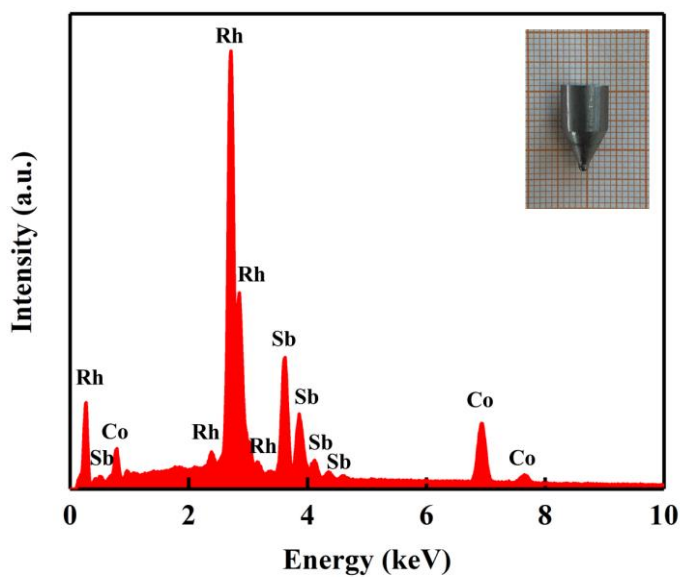

**Fig. S1. EDX curve of $Rh_2CoSb$ single crystal. The insert is the image of the single crystal.**

**Table S1. WDX data of $Rh_2CoSb$ single crystal at ten different points.**

| Point | Rh(at.%) | Co(at.%) | Sb(at.%) |
|---|---|---|---|
| 1 | 50.7 | 25.2 | 24.1 |
| 2 | 50.7 | 25.4 | 24.0 |
| 3 | 50.3 | 25.7 | 24.0 |
| 4 | 49.9 | 26.1 | 24.1 |
| 5 | 50.0 | 25.9 | 24.1 |
| 6 | 50.4 | 25.6 | 24.0 |
| 7 | 50.4 | 25.6 | 24.1 |
| 8 | 50.7 | 25.4 | 24.0 |
| 9 | 50.3 | 25.5 | 24.2 |
| 10 | 50.1 | 25.8 | 24.0 |
| Average | 50.3±0.29 | 25.6±0.27 | 24.1±0.07 |

## 2. Orientation

The quality and orientation of the single crystal were confirmed by Laue method,

which shows clear spots in both *c* and *a* axes, fitting well with the simulation.

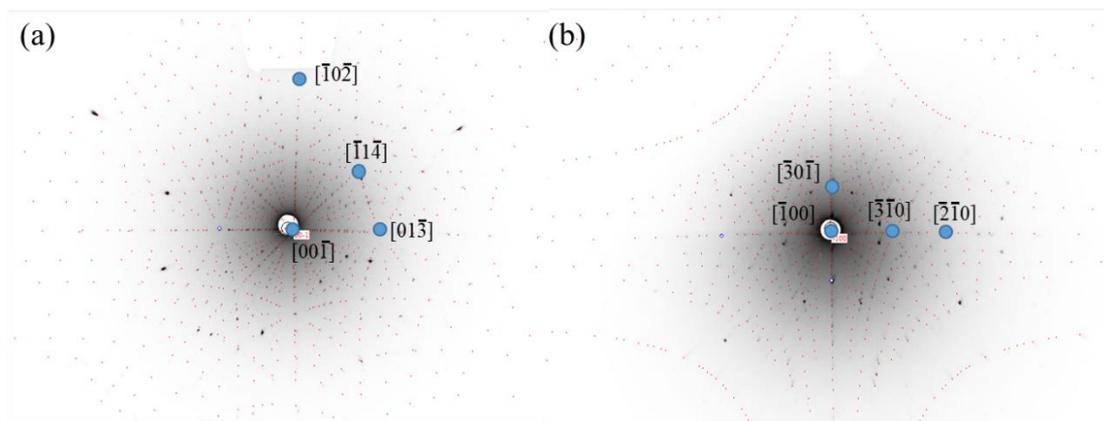

**Fig. S2. Laue pattern of Rh$_2$CoSb single crystal from *c* (a) and *a* (b) axes.**

### 3. Crystal structure

The tetragonal structure was identified by both powder X-ray diffraction and high-resolution transmission electron microscopy (TEM). The data show that Rh$_2$CoSb has a well-ordered tetragonal D0$_{22}$ structure with $a$ = 4.0393 (6) Å and $c$ = 7.1052 (7) Å. Here, the 4*d* (0 0 ¼) site is occupied by Rh, 2*b* (0 0 ½) by Co, and 2*a* (0 0 0) by Sb. The tetragonal distortion is $c/\sqrt{2}a$ = 1.2436(3).

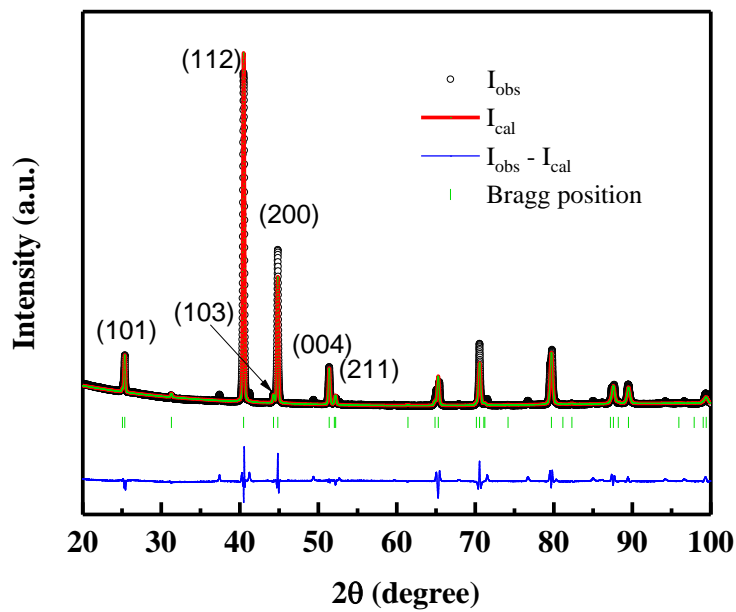

**Fig. S3. Powder X-ray diffraction of Rh$_2$CoSb at room temperature.**

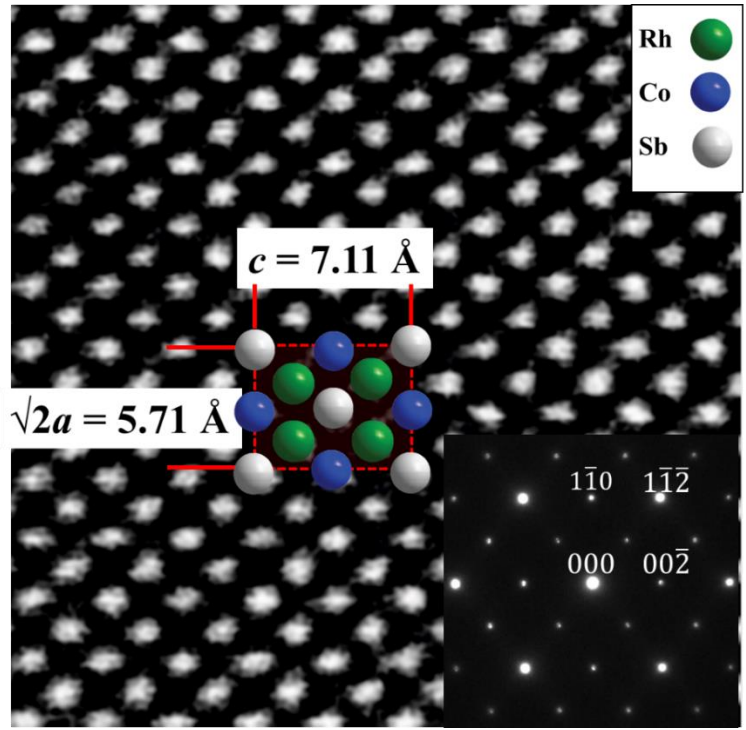

**Fig. S4. High-resolution TEM image of the [110] plane, showing a tetragonal lattice with $c/\sqrt{2}a = 1.24$.** The insert is the selected area electron diffraction image, whose superlattice reflection indicates a well-ordered structure.

## 4. Thermal analysis

Differential scanning calorimeter (DSC) and thermogravimetric (TG) analysis (see Fig. S5) have been performed to investigate the phase transitions at high temperature. Unlike many Mn-based Heusler compounds which are often cubic at high temperature and may experience a martensitic transition to become tetragonal, $Rh_2CoSb$ does not have first order transition at high temperature and is a single phase at all temperature range up to the melting point of about 1482 K, according to DSC and TG (see Fig. S5). Therefore, the common problem of twinning in many permanent magnets such as MnAl, MnGa, FePt and CoPt does not exist and single crystals have been successfully grown the by Bridgeman method.

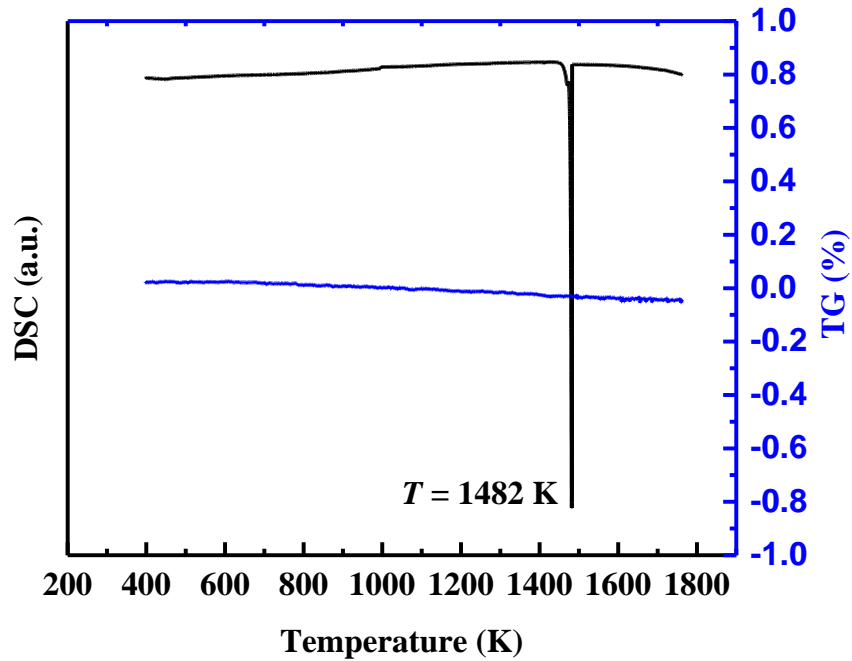

**Fig. S5. DSC and TG curve of $Rh_2CoSb$ cooling from the melt.** The TG curve (blue) is flat, indicating little Sb loss during heating. The peak in the DSC curve (black) is referred as the freezing point of 1482 K. No other peaks are detected, excluding any other first order transition. The measurement rate of thermal analysis was set as 10 K/min.

5. **Magnetization measurement**

    The single crystal magnetization curve along the *c* axis is shown in Fig. S6(a), whose enlarged part at low field is shown in (b). During the demagnetization, the sudden drop of the magnetization, namely Barkhausen jumps, together with a high susceptibility after thermal demagnetization, suggest single crystal $Rh_2CoSb$ as a nucleation type permanent magnet. The magnetization curve along the *a* axis in (c) is not saturated at 7 T, indicating strong magnetocrystalline anisotropy. $\mu_0 M_s^2/K_1$ at different temperatures is shown in (d). The peak near $T_c$ is responsible for the Hopkinson effect in Figure 2d. A small piece of single crystal was ground into powder and then annealed at 800 K for 5 h to remove the inner strain before bonded in epoxy at 400 K to make an isotropic bonded magnet. Coercivity, as an extrinsic property, is determined mainly by additional structural features such as lattice defects, grain boundaries, sample or particle size, and surface irregularities. The coercivities of bonded isotropic magnets with grain sizes of ~20 μm and 1-3 μm show a significant increase from 0.2 T to 0.9 T at room temperature. The *B-H* curve is shown in (f) for the isotropic magnet (1-3 μm grain size). The energy product $(BH)_{max}$ is about 10 kJm$^{-3}$, which is only a quarter of the theoretical value. To find a secondary phase and good alignment remain to be done for further study. For polycrystalline samples produced by arc-melting shown in (g), the coercivity is about 0.2 T.

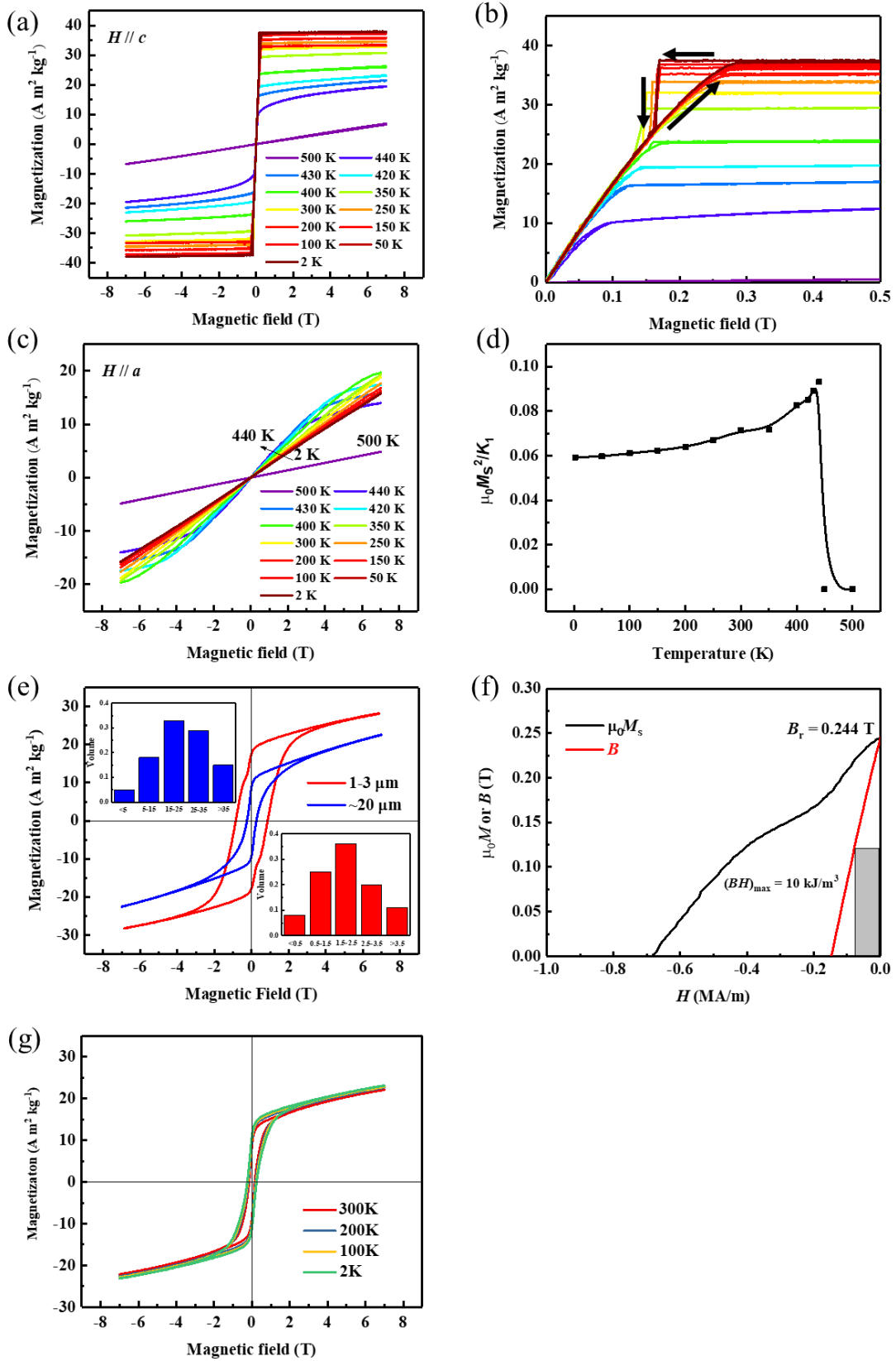

**Fig. S6. Magnetization curves of Rh$_2$CoSb.** Single crystal data from the *c* axis at different temperatures are shown in (a), whose enlarged part at low field is shown in (b). Single crystal data from the *a* axis at different temperatures are shown in (c). (d) $\mu_0 M_s^2/K_1$ at different

temperatures. *M-H* curve (e) of isotropic bonded magnets with a grain size of about 1-3 μm (red) and ~20 μm (blue) at 300 K. The insert shows their distribution histogram (unit: μm). The *B-H* curve of the magnet with grain size of 1-3 μm at 300 K is shown in (f). Magnetization curve of polycrystalline sample is shown in (g).

### 6. Electrical Transport measurement

Transport properties including the resistivity and Hall effect of $Rh_2CoSb$ are shown in Fig. S7 and S8. The resistivity is very anisotropic. With increasing temperature, the *a*-axis resistivity increases monotonically from 53 μΩ cm at 2 K to 192 μΩ cm at $T_c$, after which the resistivity is then dominated by spin disorder scattering and increases slowly[1]. The resistivity above $T_c$, which is close to the minimum metallic conductivity (~200 μΩ cm)$^{-1}$, indicates that the mean free path is close to the interatomic spacing when the moments are disordered. The *c* axis resistivity is half as large, 21 μΩ cm at 2 K but it increases with temperature and shows a similar trend to the *a* axis resistivity. Domain walls along the *c* axis in most of the volume, but not in the surface region as shown in the insert of Fig. 2a and Supplementary Fig. S11, influence the resistivity. When the charge carriers travel along *c* the current is parallel to the moment direction; but when the carriers move in the basal plane they tend to follow helical paths due to the Lorenz force, leading to a larger resistivity[2]. Like other ferromagnets, $Rh_2CoSb$ shows a negative transverse magnetoresistance (MR), which increases with increasing temperature, reaching -1.75 % at 300 K under a 7 T field. However, when the field is applied along *a* (taking care to immobilize the crystal) and the current is passed in a perpendicular *b* direction, the MR is positive (0.39%) at 300 K during the hard axis magnetization process, as shown in Fig. S7c.

The results of the Hall measurements are shown in Fig. S7d. When the field is parallel to the *c* axis and the current is along *a*, the normal Hall effect due to the Lorenz force is positive and increases linearly with applied field, indicating mainly hole-type charge carriers with a density of $1.5 \times 10^{22}$ cm$^{-3}$ corresponding to 0.86 per formula. The mobility is 2-7 cm$^2$V$^{-1}$s$^{-1}$ between 2-300 K. Detailed data can be seen in Supplementary Fig. S8. The Hall effect was also measured with the field along the *a* axis and the current along *b* at 300 K. The magnetization at a field of 9 T is far from saturation, hence the data are a mixture of a normal and an anomalous Hall effect. Their values could be estimated by linear extrapolation of the curve to 12.4 T, the saturation field from Fig. 2a, and extrapolation from the normal Hall effect. The significantly larger anomalous Hall resistivity of 1.33 μΩ cm measured along the *a* axis than the value of 0.75 μΩ cm measured along the *c* axis is attributed to the anisotropic crystal and electrical structure.

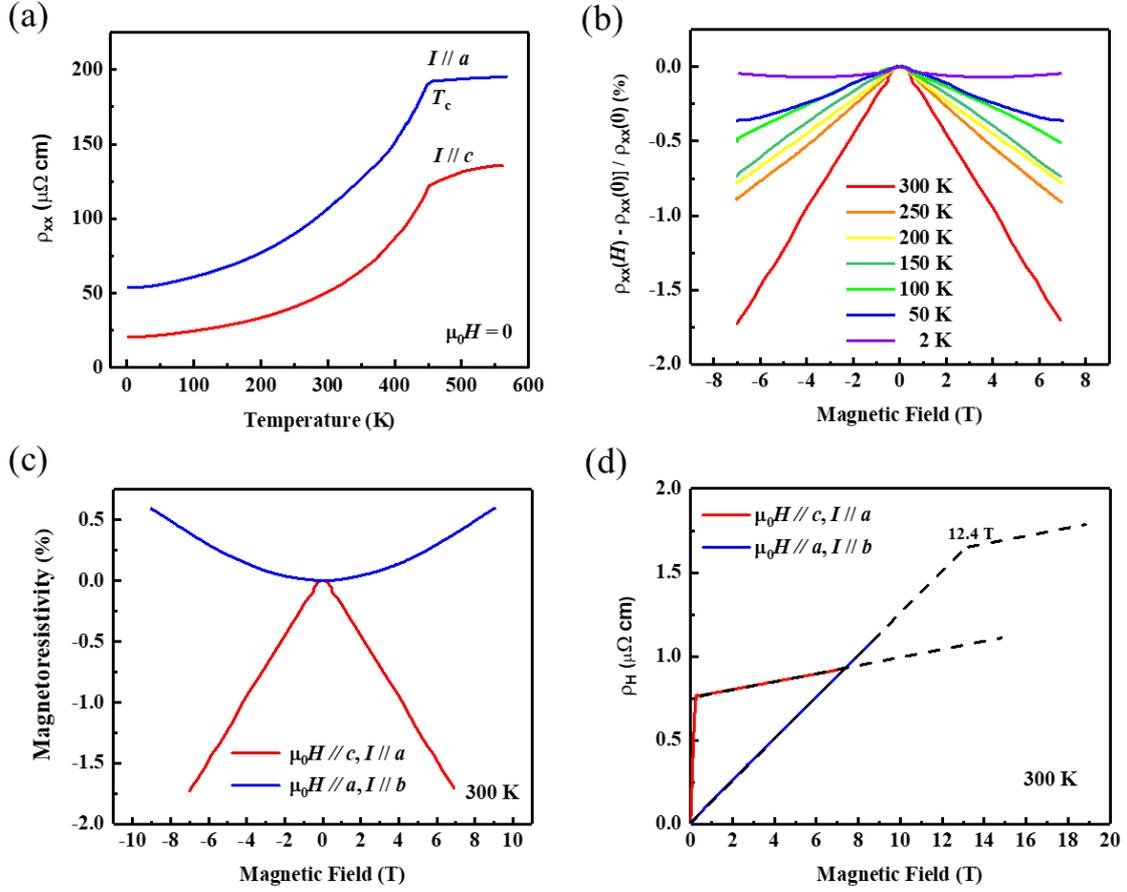

**Fig. S7. Transport properties of Rh$_2$CoSb.** a, Longitudinal resistivity along the *c* (red curve) and the *a* axis (blue curve). b, Magnetoresistance along *a* with field along *c*. c, Anisotropic magnetoresistance. d, Anomalous Hall effect with field along *a*.

The electrical transport measurement was measured both along *c* and *a* axes. The Hall conductivity was calculated by:

$$\sigma_{xy} = -\rho_{xy} / (\rho_{xy}^2 + \rho_{xx}^2)$$

where $\rho_{xx}$ and $\rho_{xy}$ are the longitudinal resistivity (along *a*) at zero field and the Hall resistivity (along *a*), respectively[3]. Since $\rho_{xx} \gg \rho_{xy}$ and the anomalous Hall resistivity $\rho_{xy}^A$ is proportional to the square of $\rho_{xx}$ as shown in the insert of Fig. S8b, the anomalous Hall conductivity is almost unchanged with variation of the temperature, indicating a side-jump effect or an intrinsic mechanism due to the Berry curvature mechanism[4] as there are a lot of band crossings in the ferromagnetic state (see Supplementary Fig. S13). The $\sigma_{xx}$ is of order $10^4$ $\Omega^{-1}$ cm$^{-1}$, which is also in the region of intrinsic AHE[3].

The anomalous Hall resistivity versus temperature is shown in Fig. S8c, while the charge carrier density as well as the mobility deduced from the normal Hall effect is shown in Fig. S8d. The charge carrier density is deduced by $n = 1/(eR_H)$, where $R_H$ is the slope of the $\rho_{xy}$. The mobility is calculated by $\mu = R_H / \rho_{xx}$.

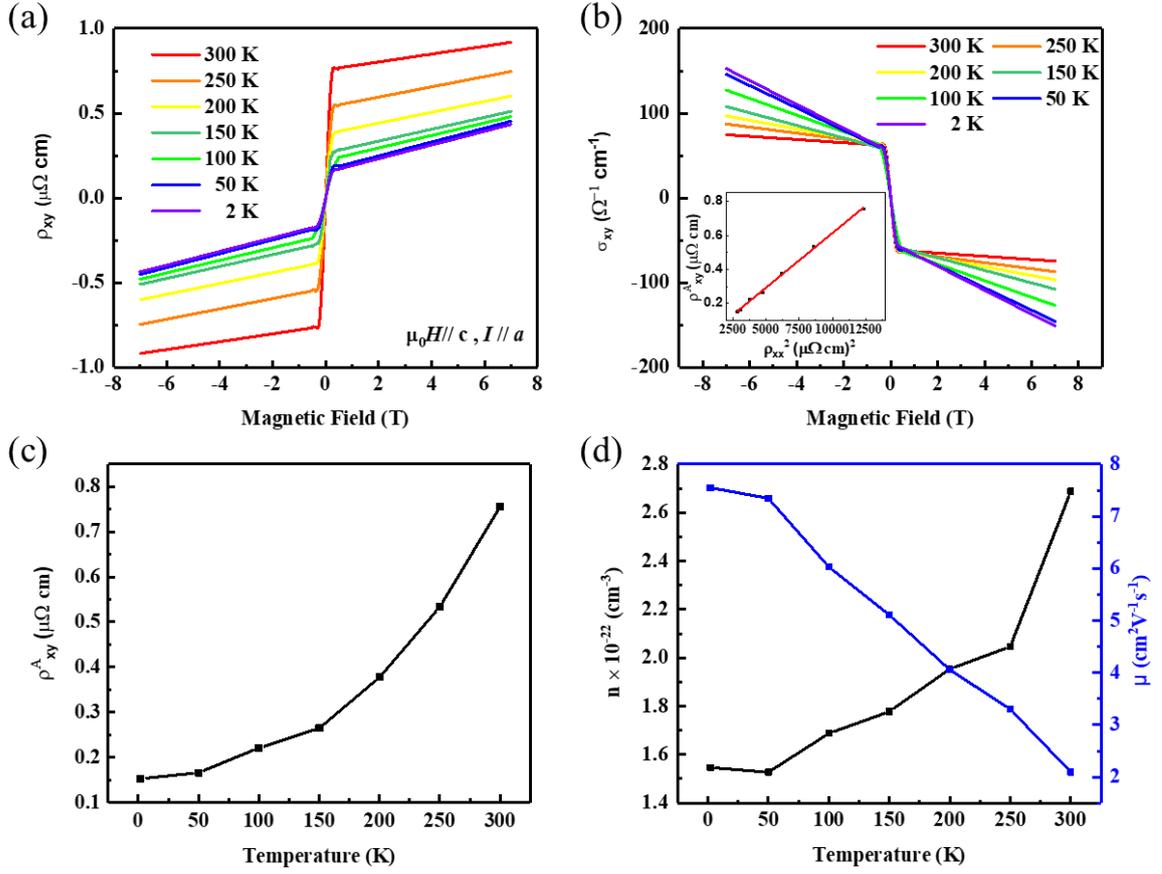

**Fig. S8.** (a) Hall effect with field along *c* at different temperatures. (b) Hall conductivity at different temperatures. The insert shows the linear fitting of anomalous Hall resistivity $\rho^A_{xy}$ versus $\rho_{xx}^2$. (c) The anomalous Hall resistivity versus temperature. (d) The charge carrier density as well as the mobility at different temperatures.

Although some quasi-two-dimensional metals[5,6] and superconductors[7-9] show a giant resistivity anisotropy (the ratio of the resistivities measured along *c* and *a* can be more than 10 to 100), the anisotropic resistivity in Rh$_2$CoSb is among the largest found in three-dimensional bulk materials. For the tetragonal Heusler compound Mn$_{1.4}$PtSn, the resistivity along *c* is only about 7% larger than along *a*[10]. Orthorhombic UFe$_2$Al$_{10}$ shows a less than 10% difference between *a*, *b* and *c*[11]. Hexagonal Mn$_3$Ge exhibits a 2.5 times larger resistivity along *a* at 2 K, but the difference decreases with increasing temperature, and vanishes around room temperature[12]. The anisotropic transport properties, including magnetoresistance and anomalous Hall effect, indicate that the magnetism leads to an anisotropic mobility of the charge carriers.

## 7. Seebeck coefficient measurement

The Seebeck coefficient was measured both along *c* and *a* axes. From Figure S9 it is seen that the absolute value of the Seebeck coefficient is slightly higher along the *c*-axis for *T* > 10 K. The behavior of the anisotropic Seebeck coefficients is reciprocal to the conductivity as expected from the Onsager relation. This reflects the anisotropy of the Onsager coefficients that are tensors. In tetragonal materials one has $S_{xx} = S_{yy} \neq S_{zz}$. It should further be noted that the elements of the Onsager tensors depend on the magnetization of the sample or the applied magnetic field, in general[13,14].

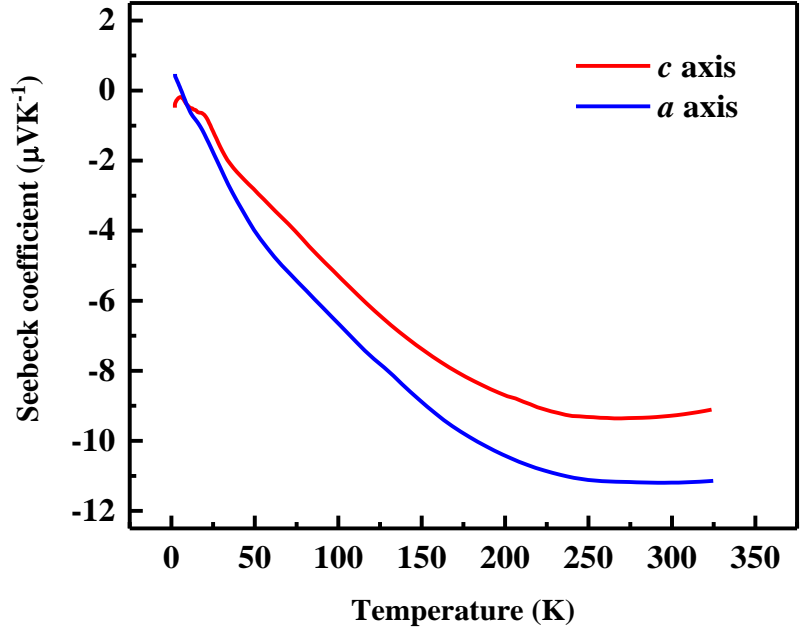

**Fig. S9. Seebeck coefficient measurement along *a* and *c* axes at different temperatures.**

## 8. Spin polarization

The spin polarization *P* of the electrons at the Femi level was deduced from point contact Andreev reflection measurement at 2 K that is shown in Fig S10. Data are analysed using the modified BTK model, as described in detail elsewhere[15]. The best fit to the spectrum is obtained with a barrier parameter *Z~* of 0.35, an electron temperature of 2 K and a spin polarization of 13%.

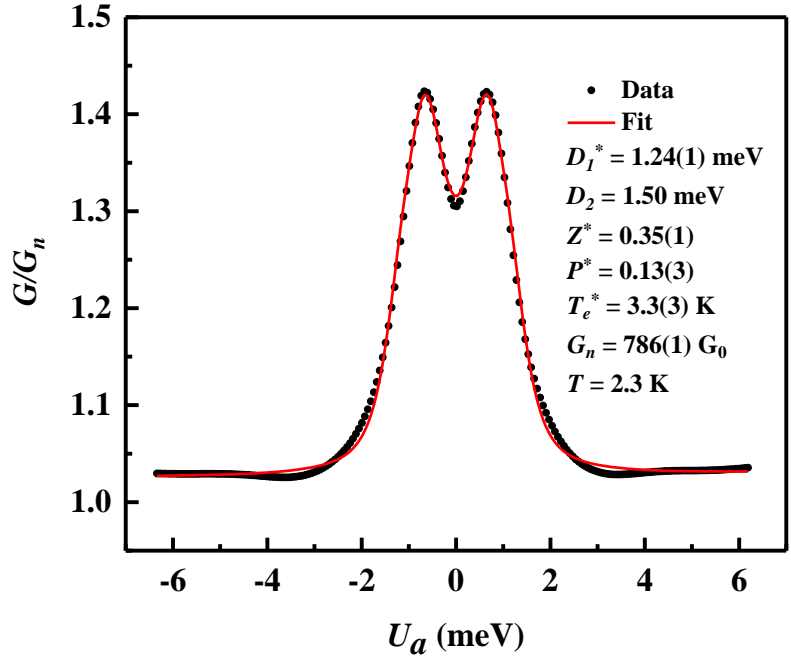

**Fig. S10.** The spin polarization $P$ of the electrons at the Femi level was deduced from point contact Andreev refection measurement at 2 K. The fit parameters for the BTK model are included in the figure (see also Ref.[15])

### 9. Magneto-optical Kerr microscopy

Magneto-optical Kerr microscopy study shows the surface domains of a two-phase branched domain pattern of higher generation on the (001) surface, which is usually observed in uniaxial magnets. For such a pattern, the domain walls are aligned parallel to the $c$-axis only in the volume of the bulk crystal. When approaching the surface, the magnetization strictly follows the $c$-axis, but the domain walls does not follow, as shown in Fig. S11. Further explanation can be found in Ref. 16.

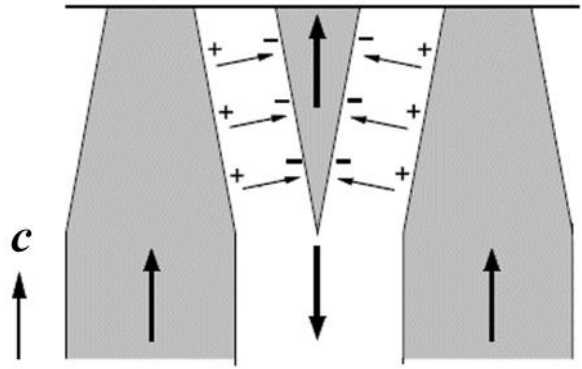

**Fig. S11.** The surface domain pattern of a strong uniaxial magnet adapted from Ref. 16.

### 10. Electron structure calculations

1) **Ab-initio calculations.**

The electronic and magnetic structures of $Rh_2CoSb$ were calculated by means of the first principles computer programs Wien2k[17-20] and SPKKR[21,22] in the local spin density approximation. In particular, the generalized gradient approximation (GGA) of Perdew, Burke and Ernzerhof[23] was used for the parametrization of the exchange correlation functional. A k-mesh based on 126 × 126 × 126 points of the full Brillouin zone was used for integration when calculating the total energies for determination of the magnetocrystalline anisotropy. For more details see [arxiv.]

2) **Results.**

The electronic and magnetic structure calculations confirm the uniaxial magnetocrystalline anisotropy with $K_u \approx 1.4$ MJ/m³ A total moment of 2.19 μB is found from the GGA calculations using SPRKKR, the spin and orbital moments have values of 2.04 $\mu_B$ and 0.15 $\mu_B$, respectively.

From the spin and site resolved data, it is found that Co contributes per atom a magnetic moment of 1.81 $\mu_B$ of which about 0.14 $\mu_B$ is the orbital moment, and the major part of 1.67 $\mu_B$ is the spin moment. A small spin moment of about 0.2 $\mu_B$ and negligible orbit moment is carried by Rh.

Both, anisotropy constant and magnetic moments are smaller than the experimental values. Indeed, $K_u$ does not include any temperature or macroscopic magnetic effects e.g.: domains and domain walls. The orbital magnetic moment of Co is larger in the XMCD measurements. However, these measurements observe an excited state. Using the orbital corrected potential of Brooks[24], the value is increased to about 0.4 $\mu_B$ without changing the spin moment.

Figures S12 and S13 show results of the ab-initio calculations using SPRKKR. Figure S12 shows the spin and site resolved density of states and Figure S13 the spin resolved band structure.

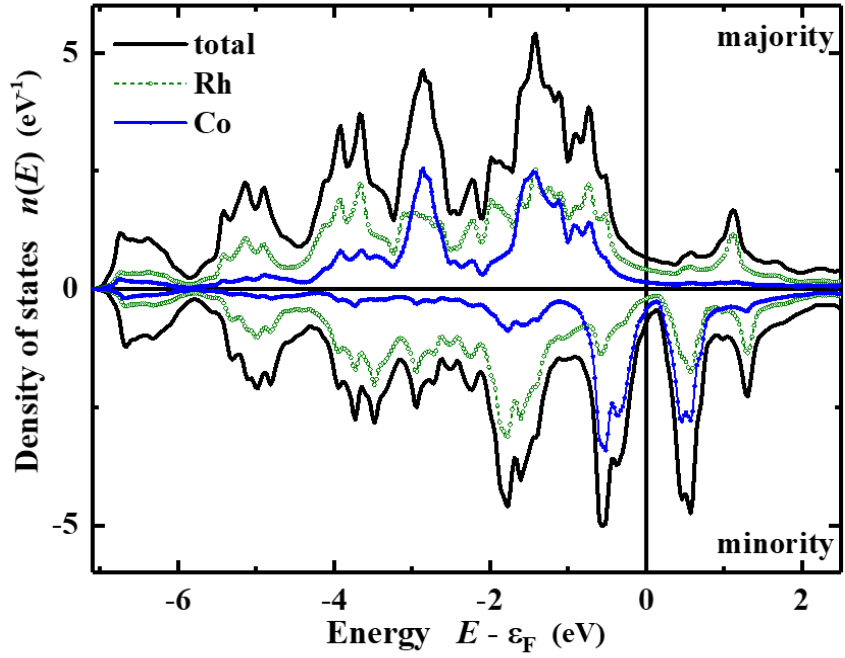

**Fig. S12. The valence band density of states.**

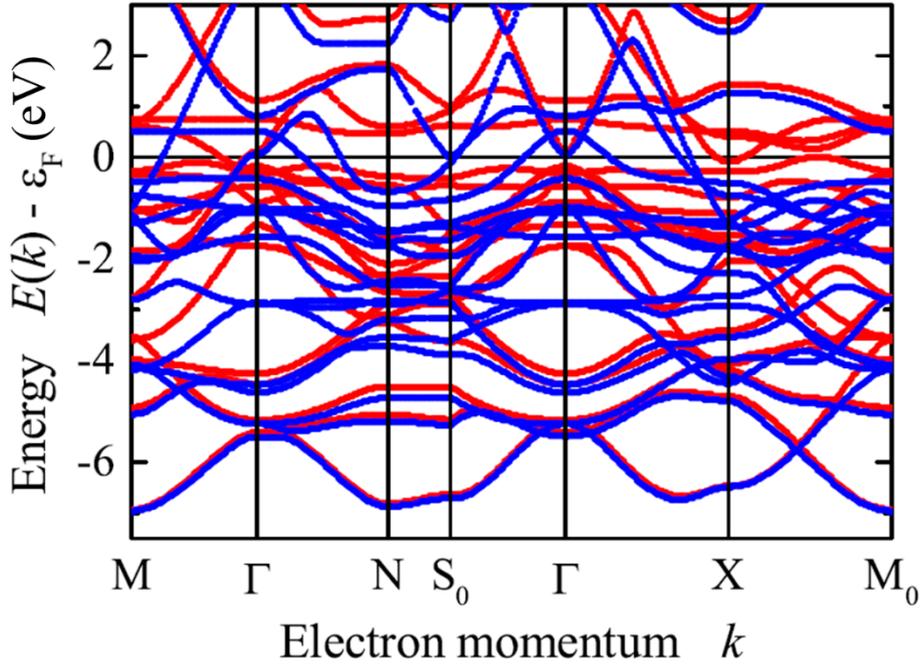

**Fig. S13. Semi-relativistic band structures of Rh$_2$CoSb. Majority states are drawn in red and minority states in blue.**

3) **Comparison of Rh$_2$FeSb and Rh$_2$CoSb**

Table S2 shows the calculated occupation of the 3$d$ orbitals of the Co and Fe atoms (number of electrons in a sphere within muffin tin radius) in Rh$_2$FeSb and Rh$_2$CoSb. Part of the $d$-electron density is delocalized and found in between the atoms and part is more localized on Rh. At both atoms the five majority-spin *d*-states are almost fully occupied (about 4.5 out of 5) resulting in a nearly spherical distribution. The occupation of the minority-spin states is quite different. In particular, the occupation of the minority $d_{z^2}$ orbital of Co modifies the spin of the charge density of Co compared to Fe. The different occupancy of the minority orbitals is responsible for the difference in anisotropy—easy plane for Fe and easy axis for Co. The calculations were performed with Wien2k.

**Table S2. Occupation of the orbitals at the Fe and Co atoms in Rh$_2$FeSb and Rh$_2$CoSb.**

| Point | Rh$_2$FeSb | | Rh$_2$CoSb | |
|---|---|---|---|---|
| | Majority | Minority | Majority | Minority |
| $d_{z^2}$ | 0.946 | 0.106 | 0.934 | 0.862 |
| $d_{xy}$ | 0.915 | 0.538 | 0.909 | 0.823 |
| $d_{x^2-y^2}$ | 0.917 | 0.271 | 0.912 | 0.426 |
| $d_{xz}$ | 0.910 | 0.377 | 0.899 | 0.385 |
| $d_{yz}$ | 0.910 | 0.377 | 0.899 | 0.385 |
| total | 4.594 | 1.668 | 4.553 | 2.881 |
| moment | 2.926 | | 1.672 | |

The Co atoms have a rather large orbital magnetic moment in Rh$_2$CoSb, from both experiments and *ab-initio* calculations. The calculated value for Co in Rh$_2$CoSb was found to be about 8% of the spin moment, whereas it is 4.8% in elemental α-Co (hcp) and 3% for Fe in the sister compound Rh$_2$FeSb. The larger orbital moments for Co reflect hybridization of Co 3*d* bands with 4*d* bands of Rh, which has a strong spin-orbit interaction.

## 11. Grain size and data storage density

The superparamagnetic blocking radius[25] for a particle is calculated as $R_b = (6k_B T/K_1)^{1/3}$, where $k_B$ and $T$ are Boltzmann constant and temperature respectively. For Rh$_2$CoSb, it is $R_b = 2.3$ nm at 300 K. For modern perpendicular recording in Co-Cr-Pt, the magnetic media are not particles, but tall slim grains. The aspect ratio is about 3, in order to increase the shape anisotropy. For hard magnets like FePt and Rh$_2$CoSb one does not need to further increase the anisotropy by larger aspect ratios. However, it is difficult to control a uniform grain height in short grains in order to get a stable Curie temperature. As a result, the ideal aspect ratio is about 1.5 to 2[26]. To resist demagnetization by random thermal fluctuation, the volume *V* of the magnetic material must satisfy the empirical condition that $K_1 V/k_B T > 60$. Therefore, *V* must be larger than 113 nm$^3$ for each grain and the diameter is calculated as 4.2 nm for Rh$_2$CoSb at 300 K for an aspect ratio of 2. In-plane heat transfer is also another thing we must pay attention to. To avoid the heating by a neighbouring grain during writing as well as to increase the signal-to-noise ratio, it is better to slightly increase the centre-to-centre distance of grains. This increased distance for Rh$_2$CoSb can still be smaller than for FePt, due to the much lower Curie temperature (450 K vs 750 K) that allows to use lower temperatures. A potential storage density of more than 10 Terabit/inch$^2$ can be realized[27] when assuming that the centre-to-centre distance of grains is 5 nm with half area occupied.

## 12. Writing speed estimation

Rh$_2$CoSb has a mass density of $\rho = 11\times10^3$ kg m$^{-3}$ and its formula weight is 386.5 g mol$^{-1}$. We calculate the heat capacity to be $C_p = 4\times3R=100$ J mol$^{-1}$K$^{-1}$ or a volume heat capacity of $C_v = 2.85\times10^6$ J m$^{-3}$K$^{-1}$. The time for each grain to cool down is calculated to be $\tau_V = C_v\cdot A/\lambda$. That is $\tau_V = 3.56\times10^{-12}$ s, when using a thermal conductivity of $\lambda = 20$ Wm$^{-1}$K$^{-1}$ and assuming an area for each grain (5 nm in diameter) of $A = 2.5\times10^{-17}$m$^2$,

The angular velocity of a hard disk is $\omega = 7000$ r min$^{-1}$ and the radius for a one-inch disk is $r = 1$ inch $= 0.0254$ m. Therefore, one has a scanning velocity of $v = \omega\cdot2\pi r = 18.56$ m s$^{-1}$. The time of heating for each grain is $\tau_h = \sqrt{A}/v = 2.7\times10^{-10}$ s and the number of heated grains becomes $n_{grains}= \tau_h/\tau_V = 75.8$.

To complete a magnetization process,

$$\frac{dH_c}{dT}\cdot\frac{dT}{dt}\cdot\frac{dt}{dx}\cdot\frac{\Delta x}{\tau_V}\cdot\frac{1}{f_{FMR}} = \Delta H_c$$

where $H_c$ is the coercivity, $T$ is temperature, $t$ is time, $x$ is the position that satisfies $dx/dt = V$ and $\Delta x =\sqrt{A}$, $f_{FMR}$ is the ferromagnetic resonance frequency (for large angle precession) at the temperature point of the maximal thermal gradient. It is a measure of how fast the switching proceeds within a grain in the trailing edge. No critical differences in $f_{FMR}$ are expected between the materials. The field of the write head $H_w$ should satisfy $\frac{H_w}{H_c-\Delta H_c} > 1$. The writing speed is proportional to $\frac{dH_c}{dT}\cdot\frac{1}{\tau_V}$.

Similarly one has for FePt $C_v = 3.1\times10^6$ J m$^{-3}$K$^{-1}$ and gets $\tau_V = C_v\cdot A/\lambda = 7.0\times10^{-12}$ s, which is twice as large as the value for Rh$_2$CoSb. Though it is unknown for the coercivity in films for Rh$_2$CoSb, we assume that $H_c$ is proportional to the anisotropy field $H_a$, hence $\frac{dH_c}{dT}$ is proportional to $(H_{a(300K)}-0)/(T_c-T_{300K})$. The $H_a$ for Rh$_2$CoSb and FePt are 12.4 T and 11.0 T at room temperature, and their $T_c$ are 450 K and 750 K, respectively. Therefore, $\frac{dH_c}{dT}$ for Rh$_2$CoSb is about 3.4 times larger than that for FePt. The writing speed ratio for Rh$_2$CoSb and FePt, proportional to $\frac{dH_c}{dT} \cdot \frac{1}{\tau_V}$, is therefore 6.8.